\documentclass{aa}  

\usepackage[utf8]{inputenc}
\usepackage{natbib,twoopt}
\usepackage[breaklinks=true]{hyperref} 
\bibpunct{(}{)}{;}{a}{}{,}             
\usepackage{graphicx}
\usepackage{txfonts}
\usepackage{hyperref}
\hypersetup{
    colorlinks=true,
    linkcolor=blue,
    citecolor = blue,
    filecolor=magenta,      
    urlcolor=blue}

\defcitealias{cerny_26}{C26}

\authorrunning{Alvarez Garay et al.}
\titlerunning{---}

\begin{document}

\title{Detailed chemo-dynamical characterization of four elusive Milky Way star clusters}

\author{D. A. Alvarez Garay\inst{1}, E. Ceccarelli\inst{2,3,4}, D. Massari\inst{3}, M. Bellazzini\inst{3}, A. Mucciarelli\inst{4}\fnmsep\inst{3}, M. Libralato\inst{5}, E. Pancino\inst{1}, A. Avdeeva\inst{1}, M. Echeveste\inst{1}, E. I. Leitinger\inst{1}\fnmsep\inst{4}, S. Nedhath\inst{1}\fnmsep\inst{6}, S. Rani\inst{1}, E. Reggiani\inst{1}\fnmsep\inst{6}, N. Sanna\inst{1}, S. Saracino\inst{1}, L. Steinbauer\inst{1}\fnmsep\inst{6}, A. Turchi\inst{1}, E. Dodd\inst{7}, E. R. Garro\inst{8} and C. Zerbinati\inst{3}
}

\institute{INAF - Osservatorio Astrofisico di Arcetri, Largo E. Fermi 5, 50125 Firenze, Italy\\
   \email{deimer.alvarez@inaf.it}
\and
Kapteyn Astronomical Institute, University of Groningen, Landleven 12, 9747 AD Groningen, The Netherlands
\and
INAF, Osservatorio di Astrofisica e Scienza dello Spazio di Bologna, Via Gobetti 93/3, I-40129 Bologna, Italy
\and
Dipartimento di Fisica e Astronomia “Augusto Righi”, Alma Mater Studiorum, Università  di Bologna, 
Via Gobetti 93/2, I-40129 Bologna, Italy
\and
INAF - Osservatorio Astronomico di Padova, Vicolo dell'Osservatorio 5, Padova 35122, Italy
\and
Dipartimento di Fisica e Astronomia, Università di Firenze, Via G. Sansone 1, 50019 Sesto Fiorentino, FI, Italy
\and
Institute for Computational Cosmology \& Centre for Extragalactic Astronomy, Department of Physics, Durham University, South Road, Durham, DH1 3LE, UK
\and
ESO – European Southern Observatory, Alonso de Cordova 3107, Vitacura, Casilla 19001, Santiago, Chile
}
 
\abstract
{A variety of faint and observationally challenging stellar systems has been discovered in the Milky Way in the last decades.
Unveiling their kinematic and chemical properties is crucial to understand their nature and to assess their role in the history of our Galaxy. 
In this paper, we present a chemo-dynamical analysis of four elusive stellar systems (BH~140, Garro~01, BLISS~1, and Kim~3) using spectra collected with the X-shooter medium-resolution spectrograph at the VLT. We derived precise line-of-sight velocities and chemical abundances for all targets. Specifically, we provide the first spectroscopic characterization of metallicity and $\alpha$-element content for BH~140, the first comprehensive chemical characterization for BLISS~1 and Garro~01, and the first spectroscopic metallicity determination for Kim~3. 
BH~140, Garro~01  and BLISS~1 are relatively metal-rich systems with $\rm [Fe/H]\sim-0.7~dex$ and have enhanced $\alpha$-element abundances, whereas Kim~3 is notably metal-poor with $\rm [Fe/H]\sim-1.8~dex$. 
We also performed an orbital analysis to investigate their origin as accreted or {\it in situ} systems.
We found that Garro~01 and BH~140 have orbits confined within $\le 1.5$~kpc from the Galactic plane, the former having also a remarkably circular orbit. They were likely born {\it in situ}, although the high orbital eccentricity of BH~140 and its mild deficiency in $\alpha$-elements leave other possibilities open. Although undoubtedly accreted, BLISS~1 does not exhibit a clear dynamical association with any known progenitor, lying in a high energy (E) - high vertical angular momentum (L$_Z$) prograde orbit reaching a maximum height from the Galactic plane of $\rm z_{max}\simeq 8~kpc$. Its chemical properties seem to favor an {\it in situ} origin, which might be connected with the Aleph sub-structure.
Finally, we tentatively link the high energy and low L$_Z$ orbit of Kim~3 to the Elqui stream.
}

   \maketitle
%
\section{Introduction} \label{sect:intro}
The advent of large-scale photometric and astrometric surveys, such as the Sloan Digital Sky Survey (SDSS; \citealt{york_00}), the Dark Energy Survey (DES; \citealt{des_05}), the Two Micron All Sky Survey (2MASS; \citealt{skrutskie_06}), the VISTA Variables in the Vía Láctea and its eXtension (VVV, \citealt{minniti_10}; VVVX, \citealt{saito_24}), and the {\it Gaia} mission \citep{gaia_16}, has revolutionized our census of Galactic star clusters \citep[e.g.,][]{garro_24}. 
In recent years, these surveys have led to the discovery of numerous faint or highly obscured stellar systems residing in both the Galactic halo and the disk/bulge \citep[e.g.,][and references therein]{balbinot_13,kim_15,kim_16,koposov_17,torrealba_19,mau_19,garro_20,garro_22,garro_22b}.

\begin{figure}[h]
\centering
\includegraphics[width=9cm]{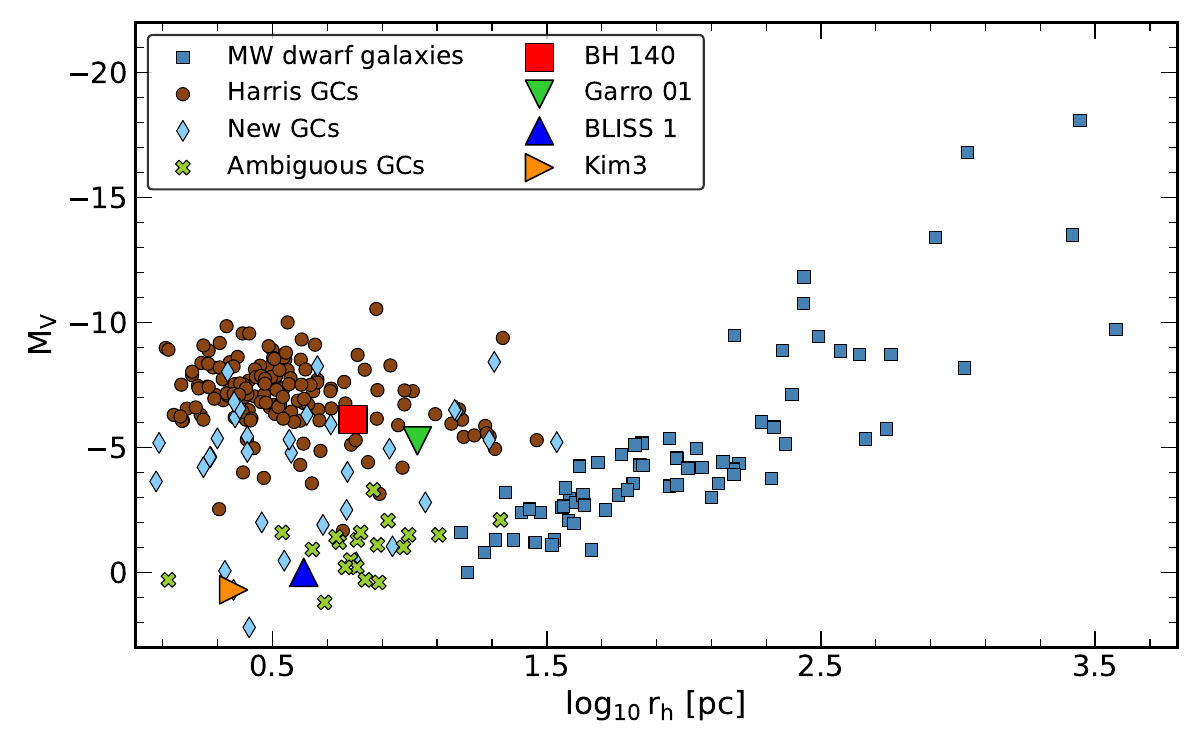}
\caption{Absolute integrated magnitude as a function of the half-light radius (in $\log$ scale) for BH~140 (red square), Garro~01 (green triangle), BLISS~1 (blue triangle), and Kim~3 (orange triangle). As a reference (with different symbols and colors) are also plotted the MW dwarf galaxies (blue squares), the GCs from the Harris catalog (brown circles), the systems classified as new GCs (light blue diamonds), and the systems whose classification is ambiguous (light green crosses). All the data come from the local volume database by \citet{pace_25}.
} \label{fig:rh_mv}
\end{figure}

The nature of a significant fraction of these new systems remains ambiguous. Indeed, many of these ultra-faint stellar systems are rapidly filling the gap between globular clusters (GCs) and ultra-faint dwarf (UFD) galaxies in the size–luminosity plane (see Fig.~\ref{fig:rh_mv}), making this diagnostic diagram less effective in determining their true physical nature (\citealt{gilmore_07,belokurov_14,laevens_14,contenta_17}; \citealt[hereafter \citetalias{cerny_26}]{cerny_26}).
To properly classify these faint systems, detailed chemo-dynamical characterization is essential. Indeed, the fundamental distinction between a UFD galaxy and a GC relies on the presence of a dark matter halo and variations in the iron content. The UFD galaxies are dark matter dominated systems; consequently, they exhibit dynamically resolved velocity dispersions that are too high to be explained by their visible stellar mass alone, leading to extremely high mass-to-light ratios \citep[e.g.,][]{simon_19}. Conversely, GCs are purely baryonic systems \citep[see,][]{gratton_19}. Chemically, the deep potential well of UFD galaxies allowed them to retain supernova ejecta and sustain multiple bursts of star formation, resulting in a measurable, intrinsic spread in their iron content ($\rm [Fe/H]$) \citep{willman_12}. In contrast, GCs generally display a homogeneous iron abundance.
Furthermore, measuring precise and detailed abundance patterns serve as key discriminants to further classify these faint stellar systems into open clusters (OCs) and GCs. Indeed, GCs are traditionally more metal-poor than OCs \citep[e.g.,][]{carretta_09,spina_21,dal_ponte_25}, and are characterized by the presence of multiple stellar populations exhibiting characteristic chemical inhomogeneities in light-elements, the most widespread example being the Na-O anticorrelation \citep{carretta_09}, a feature generally absent in OCs.
It has been proposed that many of these faint clusters, particularly those residing in the Galactic halo, were accreted into the MW following the disruption of their host satellite galaxies \citep{searle_78,koposov_07,leaman_13}.
Determining their birth environment can be key to untangle the ambiguous nature of these systems. 

Therefore, to fully understand their nature (OCs, GCs, UFD galaxies) and origin as {\it in situ} or accreted systems, a detailed characterization of their chemical and dynamical properties is needed. In this context, the {\it Hubble} Missing Globular Cluster Survey (MGCS; \citealt{massari_25}) aims to shed light on these newly discovered objects. Specifically, the MGCS provides homogeneous two-band photometry and astrometry \citep{libralato_26} for all those kinematically confirmed clusters that were never observed by {\it Hubble} before. 
One of the main goals of this new survey is to determine the age and the distance of these systems, which will be combined with their chemo-dynamical properties in order to robustly assess their nature and origin \citep{massari_23}.

In this study we focus on four recently discovered systems, namely BH~140, Garro~01, BLISS~1 and Kim~3. 
BH~140 is a stellar system located at a heliocentric distance of $\rm D_{\odot} = 4.81\pm0.25~kpc$ \citep{baumgardt_21}, with a luminosity\footnote{All the luminosities and absolute magnitudes reported in this work refer to the visual $V$-band.} of $\rm L \sim 2.4\cdot10^4~L_{\odot}$ \citep{baumgardt_20}, and a mass of $\rm 6.1\cdot10^4~M_{\odot}$ \citep{baumgardt_18}.
This system is a candidate cluster for which virtually no chemical information is available in the literature. In the past it was classified as an open cluster \citep{Kharchenko_13}, but in the following years \citet{cantat_18} provided evidence\footnote{\citet{cantat_18} found that the color-magnitude diagram of BH~140 shows the typical aspect of GC, with a prominent giant branch and an interrupted horizontal branch displaying a gap in the locus occupied by RR Lyrae.} that it could be actually a GC. 
Garro~01 was recently identified by \citet{garro_20} combining VVV near-infrared photometry and {\it Gaia} DR2 proper motions. This system is located at a heliocentric distance of $\rm D_{\odot} = 15.5\pm1.0~kpc$ and from a photometric study, \citet{garro_20} classified Garro~01 as a low-luminosity ($\rm L \sim 1.09\cdot10^4~L_{\odot}$) and metal-rich ($\rm [Fe/H] = -0.7$ dex) GC with a mass of $\rm \sim10^4~M_{\odot}$, located at a low latitude in the MW disk. \citet{garro_20} estimated and age of $\rm 11.0\pm1.0~Gyr$ for this system. 
However, the true nature of Garro~01 has recently been questioned by \citet{pace_23}, who suggested that Garro~01 could be actually an open cluster. 
BLISS~1 is a stellar system recently discovered by \citet{mau_19} using multi-band imaging data from the Dark Energy Camera (DECam) as part of the Blanco Imaging of the Southern Sky (BLISS) survey. This system is located at a heliocentric distance of $\rm D_{\odot} = 23.7^{+1.9}_{-1.0}~kpc$, is faint ($\rm L \sim 85.5~L_{\odot}$) and very low-massive ($\rm M = 143_{-39}^{+34}~M_{\odot}$, \citealt{mau_19}). Based on photometry, \citet{mau_19} suggested it to be a relatively metal-poor ($\rm [Fe/H] = -1.4$ dex) system associated with the Galactic halo. BLISS~1 appears to belong to relatively metal-poor systems that resemble GCs, but at a fainter scales. Moreover, it is more compact than UFD galaxies \citep{torrealba_18}. Due to these properties, \citet{mau_19} suggested that BLISS~1 could be an ultra-faint halo star cluster rather than an UFD galaxy. 
Finally, we target Kim~3, an ultra-faint star cluster discovered by \citet{kim_16} using DECam photometry. Located at a heliocentric distance of $\rm D_{\odot}=15.14^{+1.00}_{-0.28}~kpc$, this system is characterized by an extremely low luminosity ($\rm L \sim 44.9~L_{\odot}$) and low-mass ($\rm M = 90_{-20}^{+30}~M_{\odot}$, \citetalias{cerny_26}). A previous photometric study tentatively suggested a relatively young age (9.5~Gyr) and low metallicity ($\rm [Fe/H] = -1.6~dex$, \citealt{kim_16}). 
Recently, \citetalias{cerny_26} inferred an age of $\rm \sim9.5~Gyr$ from the main sequence turn-off morphology, although without deriving a spectroscopic metallicity. Therefore, while a direct metallicity measurement is needed to confirm its nature, the system appears to be younger than the known population of MW UFD galaxies, suggesting that Kim~3 could be a faint star cluster similar to BLISS~1.

The location of the four stellar systems in the diagram opposing the $\rm r_h$ to the $\rm M_V$, shown in Fig.~\ref{fig:rh_mv}, suggest that BH~140 and Garro~01 may lie along the wide tail of the locus of "classic" GCs, slightly fainter than the peak of the distribution ($\rm M_V \sim -7.5$). 
On the other hand, BLISS~1 and Kim~3 occupy the region populated by stellar systems whose nature is not clearly established. Indeed, although these systems exhibit physical sizes consistent with the population of classical GCs (a few parsecs), they are hundreds of times fainter and broadly consistent with the extrapolation of the locus of dwarf galaxies to very faint magnitudes (see \citetalias{cerny_26}, for a thorough discussion).

We aim to obtain a complete characterization of all these new intriguing systems in terms of line-of-sight velocities ($\rm V_{los}$), metallicity ([Fe/H]), [$\alpha$/Fe] as well as other chemical elements \citep[e.g.,][\citetalias{cerny_26}] {pace_23,ceccarelli_26,petralia_26}.
Obtaining such precise chemo-dynamical information serves to trace the nature and origin of these ambiguous targets. First, searching for an intrinsic spread in metallicity is the ultimate test to distinguish an UFD galaxy from a faint star cluster \citep{willman_12,frebel_14}. 
Moreover, precise metallicities and abundance patterns can discriminate between an open and a globular cluster systems, as the latter are traditionally more metal-poor than their open cluster counterparts \citep[e.g.,][]{spina_21,dal_ponte_25,carretta_09}; 
this is particularly relevant for targets like Garro~01, whose classification is still under debate \citep[e.g.,][]{garro_20,pace_23}.
Second, by combining chemical tagging with orbital analysis, we can reconstruct their formation history and determine whether they formed {\it in situ} (definitely excluding an UFD galaxy nature) or were accreted from external galactic building blocks
\citep[see, e.g.,][]{massari_19,horta_20,monty_23,ceccarelli_24}.

This paper is organized as follows: In Sect.~\ref{sect:obs} we detail the observations and the target selection, in Sect.~\ref{sect:params} we describe the stellar parameters and velocities, in Sect.~\ref{sect:op} we illustrate the kinematical properties of the systems, in Sect.~\ref{sect:analysis} we detail the chemical analysis, in Sect.~\ref{sect:res_chem} we illustrate the results of the abundance analysis, and finally in Sect.~\ref{sect:disc} we carry out the discussion of the results.

\section{Observations and target selection} \label{sect:obs}
In this study we analyzed 15 stars belonging to four stellar clusters, namely BH~140, Garro~01, BLISS~1, and Kim~3,  all of them being targeted by MGCS. In particular, we analyzed spectra acquired at the Very Large Telescope UT3 (Melipal) with the VIS arm of the X-shooter spectrograph \citep{vernet_11}, covering the spectral range from $\sim5500$ to $\sim10200~\AA$ and employing a $0.7"$ slit to reach a spectral resolution of $\sim11400$. The spectra were acquired under the programme 0114.D-0288 (P.I. E. Ceccarelli). 

For BH~140, BLISS~1, and Kim~3 we selected target stars among the high proper-motion based probability members defined by \citet[][$\rm P>80\%$]{vasiliev_21}, while for Garro~01 we selected stars from the decontaminated catalog of highly-
probable cluster members from \citet{garro_20}. 
For BH~140 and Garro~01 we select four and three Red Giant Branch (RGB) stars per cluster, respectively.  
On the other hand BLISS~1 and Kim~3 lack any RGB member and we were forced to select four and three relatively bright unevolved upper main sequence stars, respectively. 
To check already at this point that our target selection is free from pathological cases like binary stars, we inspected the RUWE parameter given by the {\it Gaia} catalog. This value is expected to be around 1.0 for sources where the single-star model provides a good fit to the astrometric observations.  
Generally stars with $\rm RUWE > 1.4$ are considered as non-single or otherwise problematic for the astrometric solution \citep[][and references therein]{gaia_23}. All our targets have RUWE in the range $0.96 - 1.14$. 

The exposure times for each star were chosen according to their brightness. In the case of BH~140, each star was observed for a total of 90~s reaching a signal-to-noise ratio (SNR) per resolution element of $\sim20-30$ around the Mg line at $8806~\AA$. For Garro~01 the stars were observed for a total time ranging from 400 to 1500 s, allowing to reach a SNR ranging from 30 to 60. The stars of BLISS~1 were observed from 600 to 2500 s each, reaching a SNR of $\sim 20-30$. 
Finally, among the three stars of Kim~3, two of them were observed for a total time of 3600 s each, while the faintest one was observed for a total time of 10800~s, divided into three different observations of 3600 s each. 
This allows us to reach a SNR of $\sim 5-15$ for the three stars observed.

All spectra were reduced using the dedicated X-shooter pipeline\footnote{\url{https://www.eso.org/sci/facilities/paranal/instruments/xshooter.html}}, which includes bias subtraction, flat field correction, spectral extraction, wavelength calibration, and order merging.
To remove the contamination from the Earth's atmosphere, a telluric correction was applied to all the reduced spectra. Specifically, we divided each observed spectrum by a synthetic atmospheric transmission model computed specifically for the observing conditions using the TAPAS web-based facility \citep{bertaux_14}.
Table~\ref{tab:1} lists the observed target stars, their coordinates, photometric properties, and details of the observations. 
\begin{table*}[h]
\caption{Main information for the observed targets.}
\label{tab:1}
\centering
\renewcommand{\arraystretch}{1.5}
\small
\setlength{\tabcolsep}{5.6pt}
\begin{tabular}{cccccccccc}
\hline\hline 
ID & ID \textit{Gaia} & RA & Dec & $\rm G$ & $\rm BP$ & $\rm RP$ & $\rm E(B-V)$ & $\rm t_{exp}$ & SNR \\
 & {\it Gaia} DR3  & [deg] & [deg] & [mag] & [mag] & [mag] & [mag] & [s] & [8806~$\AA$]  \\
\hline
$\rm BH~140\_04$ & 5858110992638378240 & 193.4303013 & -67.2671979 & 13.8466 & 14.9655 & 12.7972 & 0.80 & 90 & 24 \\
$\rm BH~140\_06$ & 5858101097032521728 & 193.7860002 & -67.2119773 & 13.8916 & 14.7423 & 12.9736 & 0.61 & 90 & 30 \\
$\rm BH~140\_07$ & 5858114875288910464 & 193.2032870 & -67.2048141 & 14.0097 & 14.8966 & 13.0715 & 0.59 & 90 & 24 \\
$\rm BH~140\_09$ & 5858114772209644288 & 193.1525754 & -67.2249635 & 14.2505 & 15.0945 & 13.3386 & 0.63 & 90 & 18 \\
$\rm Garro~01\_01$ & 5851223376949522048 & 212.2378206 & -65.6070644 & 14.9140 & 16.3222 & 13.7647 & 0.58 & 400  & 60 \\
$\rm Garro~01\_02$ & 5851223686187222784 & 212.1601075 & -65.5750991 & 15.0318 & 16.5065 & 13.8580 & 0.58 & 400  & 52 \\
$\rm Garro~01\_09$ & 5851223372582981888 & 212.2428941 & -65.6045845 & 17.2641 & 18.2573 & 16.3056 & 0.58 & 1200 & 32 \\
$\rm BLISS~1\_01$ & 5380491044223652736 & 177.5412873 & -41.7646823 & 16.1873 & 16.8902 & 15.3809 & 0.12 & 600  & 30 \\
$\rm BLISS~1\_02$ & 5380485336211925760 & 177.5102783 & -41.7822662 & 16.5326 & 17.2093 & 15.7375 & 0.12 & 1200 & 35 \\
$\rm BLISS~1\_03$ & 5380485404931412480 & 177.4922810 & -41.7753383 & 17.8113 & 18.3369 & 17.1574 & 0.12 & 1500 & 29 \\
$\rm BLISS~1\_04$ & 5380491211727206784 & 177.5201849 & -41.7695053 & 18.5825 & 19.1195 & 17.8833 & 0.12 & 2500 & 22 \\
$\rm Kim~3\_01$ & 6181851604292811392 & 200.6831874 & -30.5986729 & 18.9330 & 19.3268 & 18.4383 & 0.06 & 3600  & 14 \\
$\rm Kim~3\_02$ & 6181851776091515904 & 200.6853598 & -30.5898324 & 19.2662 & 19.6157 & 18.9182 & 0.06 & 3600  & 9  \\
$\rm Kim~3\_04$ & 6181851707372021120 & 200.6985750 & -30.6011661 & 20.1498 & 20.3389 & 19.7195 & 0.06 & $3\times3600$ & 5  \\
									    
\hline 
\end{tabular}
\tablefoot{The coordinates RA and Dec from Gaia DR3 \citep{gaia_23} are in the  International
Celestial Reference System at epoch 2016. Also, the reported magnitudes are the observed ones, before the reddening correction.}
\vspace{2pt}
\end{table*}

\section{Atmospheric parameters and line-of-sight velocities} \label{sect:params}
\subsection{Atmospheric parameters} \label{sect:param_atm}
In this study we determined the atmospheric parameters using the photometric data present in the {\it Gaia} Data Release 3 \citep[DR3,][]{gaia_16,gaia_23}. 
We first derived the dereddened {\it Gaia} magnitudes, where the extinction coefficients were obtained using the reddening law by \citet{fitzpatrick_19}. 
For the most extinct cluster BH~140, which has $\rm E(B-V) = 0.70$ \citep{Kharchenko_13} we corrected magnitudes also for differential reddening applying the method described in \citet{milone_12}. 
The effective temperatures ($\rm T_{eff}$) were obtained using the $\rm (BP - RP)_0 - T_{eff}$ transformation by \citet{mucciarelli_21}. 
For the surface gravities ($\rm \log g$) we used the Stefan-Boltzmann relation, assuming our derived temperatures and a mass of $\rm 0.8~M_{\odot}$. 
We note that the mass assumption has a negligible impact on the derived abundances: a difference of $\rm 0.2~M_{\odot}$ in the mass does indeed lead to a modification in $\rm log~g$ of $\sim 0.1$, corresponding to a variation in the measured abundances of about 0.02~dex or less. 
Stellar luminosities were computed by using the dereddened G-band magnitude, the bolometric corrections by \citet{andrae_18}, and the true distance modulus $\rm (m-M)_0$ of each cluster. 
Propagating all the sources of error into the uncertainty on log g we found that the adoption of a 0.1~dex uncertainty on this parameter was an adequate choice in all the cases.
Finally, for the microturbulent velocity ($\rm v_t$) we employed the relation provided by \citet{kirby_09}, assuming in all the cases a conservative uncertainty of $\rm 0.1~km~s^{-1}$ on $\rm v_t$.
For reference, the main physical parameters of the four systems are listed in Table~\ref{tab:info_cl}.

To derive $\rm T_{eff}$ and $\log~g$ for the observed stars, we adopted $\rm E(B-V)$ and true distance modulus $\rm (m-M)_0$ values listed in Table~\ref{tab:1} and Table~\ref{tab:info_cl}, respectively.  
The uncertainties on $\rm T_{eff}$ were computed as the sum in quadrature of the errors associated with the photometric accuracy of the data, the total reddening (assuming a typical uncertainty of 10\%), and the intrinsic scatter of the $\rm (BP - RP)_0 - T_{eff}$ relation (83 K).
This procedure yields typical $\rm T_{eff}$ uncertainties of $\sim 200$~K for BH~140 and
$\sim 90-130$~K for BLISS~1. For Garro~01, the errors are $\sim 100$~K for the two brightest stars and $\sim 150$~K for the remaining one. Finally, for Kim~3, the uncertainties are of the order of $200-300$~K for the two brightest targets, increasing to $\sim700$~K for the faintest star due to the significantly large errors in its Gaia photometry.

All the atmospheric parameters, together with their uncertainties are reported in Table~\ref{tab:2}.

\begin{table*}[h]
\caption{Main physical parameters of the clusters analyzed in this study.}
\label{tab:info_cl}
\centering
\renewcommand{\arraystretch}{1.5}
\footnotesize
\setlength{\tabcolsep}{3.pt}
\begin{tabular}{ccccccccccc}
\hline\hline 
ID & $\rm (m-M)_0$ & $\rm M_V$ & mass & $\rm r_h$ & RA & Dec & pmra & pmdec & $\rm V_{los}$ & Ref \\
 & &  & [$\rm M_{_{\odot}}$]  & [pc] & [deg] & [deg] & [$\rm mas \cdot yr^{-1}$] & [$\rm mas \cdot yr^{-1}$] & [$\rm km \cdot s^{-1}$] & \\
\hline

BH~140 & $13.41\pm0.11$ & $-6.12\pm0.12$ & $\rm 6.1\cdot10^4$ & $6.22$ & 193.4729 & -67.1773 & $-14.853\pm0.004$ & $1.242\pm0.004$ & $90.5_{-1.6}^{+1.8}$ & [1,2,3,4] \\
Garro~01 & $15.93\pm0.03$ & $-5.26\pm1.00$ & $\rm \sim10^4$ & $10.67\pm2.20$ & 212.2500 & -65.6200 & $-4.431\pm0.015$ & $-1.127\pm0.015$ & $30.2^{+3.8}_{-3.7}$ & [4,5] \\
BLISS~1 & $16.87^{+0.17}_{-0.09}$ & $0.0^{+1.7}_{-0.7}$ & $\rm 143_{-39}^{+34}$ & $4.11\pm1.12$ & 177.5110 & -41.7720 & $-2.404\pm0.037$ & $0.111\pm0.051$ & $123.0\pm0.6$ & [4,6]  \\
Kim~3 & $15.90^{+0.14}_{-0.04}$ & $0.7\pm0.3$ & $\rm 90_{-20}^{+30}$ & $2.28\pm0.76$ & 200.6883 & -30.6010 & $-0.950\pm0.053$ & $3.831\pm0.030$ & $164.5_{-3.0}^{+2.7}$ & [4,7,8,9] \\

\hline 
\end{tabular}
\tablefoot{The references enumerated above are: [1] \citet{baumgardt_18}, [2] \citet{baumgardt_20}, [3] \citet{baumgardt_21}, [4] \citet{libralato_26},  [5] \citet{garro_20}, [6] \citet{mau_19}, [7] \citet{kim_16}, [8] \citet{pace_25}, and [9] \citetalias{cerny_26}. $\rm V_{los}$ are from this paper. \\
We recall that the luminosity is encoded in the $\rm M_V$ ($\rm L/L_{\odot} = 10^{0.4(M_{\odot} - M_V)}$), while the distance in the distance modulus through the relation $\rm (m-M)_0 = 5\log(d) - 5 $.}
\vspace{2pt}
\end{table*}

\begin{table}[h]
\caption{Stellar parameters for the observed targets.}
\label{tab:2}
\centering
\renewcommand{\arraystretch}{1.5}
\footnotesize
\setlength{\tabcolsep}{4.pt}
\begin{tabular}{ccccc}
\hline\hline 
ID & $\rm T_{eff}$ & $\rm \log g$ & $\rm v_t$ & $\rm V_{los}$ \\
 & [K] & [cgs] & [$\rm km~s^{-1}$] & [$\rm km~s^{-1}$] \\
\hline

$\rm BH~140\_04$ &  $4685\pm209$ & $1.44\pm0.10$ & $1.80\pm0.10$ & $90.7\pm0.8$ \\
$\rm BH~140\_06$ &  $5066\pm197$ & $1.80\pm0.10$ & $1.72\pm0.10$ & $87.9\pm0.9$ \\
$\rm BH~140\_07$ &  $4906\pm179$ & $1.79\pm0.10$ & $1.72\pm0.10$ & $93.6\pm1.4$ \\
$\rm BH~140\_09$ &  $5158\pm213$ & $1.96\pm0.10$ & $1.68\pm0.10$ & $91.0\pm1.5$ \\
$\rm Garro~01\_01$ & $3800\pm105$ & $0.55\pm0.10$ & $2.00\pm0.10$ & $20.8\pm0.9$ \\
$\rm Garro~01\_02$ & $3714\pm102$ & $0.53\pm0.10$ & $2.01\pm0.10$ & $30.3\pm0.9$ \\
$\rm Garro~01\_09$ & $4621\pm152$ & $1.97\pm0.10$ & $1.68\pm0.10$ & $30.0\pm1.1$ \\
$\rm BLISS~1\_01$ & $4469\pm95 $ & $1.52\pm0.07$ & $1.78\pm0.07$ & $123.7\pm0.5$ \\
$\rm BLISS~1\_02$ & $4529\pm96 $ & $1.69\pm0.07$ & $1.74\pm0.07$ & $122.2\pm0.3$ \\
$\rm BLISS~1\_03$ & $5071\pm114$ & $2.46\pm0.08$ & $1.56\pm0.08$ & $122.4\pm0.4$ \\
$\rm BLISS~1\_04$ & $4953\pm126$ & $2.72\pm0.08$ & $1.50\pm0.08$ & $124.5\pm0.7$ \\
$\rm Kim~3\_01$ & $5541\pm202$ & $3.54\pm0.09$ & $1.32\pm0.10$ & $165.7\pm2.1$ \\
$\rm Kim~3\_02$ & $6101\pm303$ & $3.85\pm0.11$ & $1.24\pm0.10$ & $164.5\pm2.4$ \\   
$\rm Kim~3\_04$ & $6370\pm700$ & $4.28\pm0.20$ & $1.15\pm0.10$ & $161.9\pm3.9$ \\   

\hline 
\end{tabular}
\vspace{2pt}
\end{table}


\subsection{Line-of-sight velocities} \label{sect:rv}
Line-of-sight velocities of individual stars were derived using the cross-correlation function (CCF) technique \citep[e.g.,][]{tonry_79}, as implemented in the \texttt{PyAstronomy} package\footnote{\url{https://pyastronomy.readthedocs.io/en/latest/pyaslDoc/aslDoc/crosscorr.html}}.
To maximize the precision of the cross-correlation, we generated an appropriate synthetic spectrum for each target using the \texttt{SYNTHE} code \citep{sbordone_04,kurucz_05}. These templates were computed adopting the specific atmospheric parameters ($\rm T_{eff}$, $\rm \log g$, and $\rm v_t$) derived above and were subsequently convolved with a Gaussian profile to match the observed spectral resolution of our stars.
The $\rm V_{los}$ uncertainties associated with the CCF measurements were derived through Monte Carlo simulations. For each target, we injected Poissonian noise into the appropriate synthetic spectrum to properly match the SNR of our X-shooter spectra. We then computed the CCF between the observed and the noise-injected synthetic spectra. This procedure was repeated 200 times per star, and the standard deviation of the resulting $\rm V_{los}$ distribution was adopted as the formal measurement uncertainty. The results of these measures are reported in Table~\ref{tab:2}.

For all the target clusters we determined the mean $\rm V_{los}$ and the intrinsic $\rm \sigma_{V_{los}}$ with a Bayesian 2-parameter fit, adopting a simple gaussian as a model for the distribution of line of sight velocities. We used {\tt emcee} \citep{emcee} to explore with Monte Carlo Markov Chains (MCMC) the likelihood function introduced by \citet{walker06} and widely used for this purpose in the literature. We adopt wide uniform priors for both parameters and we used 15 walkers each running for 10000 steps, then we remove the first 1000 steps as a burning phase. We take the median of the posterior distribution function (PDF) as our best estimate of the mean systemic $\rm V_{los}$ and the intervals between the median and the 16th and 84th percentiles as proxies for $\pm 1\sigma$ uncertainties (see Table~\ref{tab:sigma}).

While an adequate insight into the internal kinematics of ordinary star clusters like BH~140 or Garro~01 would require relatively large samples tracing the radial profile of the velocity dispersion (and possibly rotation), for the very faint and distant clusters of the class of Bliss~1 and Kim~3 a single global $\rm \sigma_{V_{los}}$ value is all we can hope to measure/constrain with the handful of stars for which $\rm V_{los}$ are typically available (see \citetalias{cerny_26}). In this context, also our small samples of $\rm V_{los}$ measurements may contribute to a better characterization of the cluster kinematics, especially given the sub- {$\rm km~s^{-1}$} uncertainty of our individual $\rm V_{los}$ measures for Bliss~1. For this reason,
in Table~\ref{tab:sigma} we present the values of the mode, 5th, 50th, and 95th percentile to provide a synthetic description of  the PDF of $\rm \sigma_{V_{los}}$ for these two clusters. In the following we briefly discuss the kinematics of the target clusters individually, in the context of the available literature.

BH~140: for this cluster we inferred a mean heliocentric $\rm V_{los} = 90.5^{=1.8}_{-1.6}~km~s^{-1}$. 
This value is in good agreement with those reported in the literature, like, e.g. $\rm V_{los} = 90.29\pm0.35~km~s^{-1}$, from Holger Baumgardt's  globular cluster database\footnote{\url{https://people.smp.uq.edu.au/HolgerBaumgardt/globular/}}. \citet{soubiran_18}, from a sample of four stars derived a mean $\rm V_{los}$ for BH~140 of $\rm 90.4\pm0.9~km~s^{-1}$, while,
recently \citet{petralia_26} found $\rm V_{los} = 91.72\pm1.92~km~s^{-1}$ from high-resolution IR spectroscopy of seven stars. 

Garro~01:  among the observed targets in this system, one star (Garro~01\_01) exhibits a $\rm V_{los} =20.8\pm 1.0~km~s^{-1}$. This value deviates by more than $10\sigma$ from the systemic velocity of the cluster reported by \citet[][$\rm \sigma_{V_{los}} = 0.4\pm0.3~km~s^{-1}$]{pace_23}. Therefore, this star is not a member of the cluster and was excluded from the following analysis. For the remaining two stars, we measured a mean $\rm V_{los} = 30.2^{+3.8}_{-3.7}~km~s^{-1}$, which is in agreement with the value of 
$\rm 31.0\pm0.1~km~s^{-1}$ obtained by \citet{pace_23} from the analysis of 42 high-resolution spectra.

\begin{table}[h]
\caption{Values of PDF for the velocity dispersion of Bliss~1 and Kim~3.}
\label{tab:sigma}
\centering
\renewcommand{\arraystretch}{1.5}
\footnotesize
\setlength{\tabcolsep}{4.pt}
\begin{tabular}{lccccc}
\hline\hline 
ID & $\rm P_{5\%}$ & mode & $\rm P_{50\%}$ & $\rm P_{95\%}$ & N \\
   & [$\rm km~s^{-1}$] & [$\rm km~s^{-1}$] & [$\rm km~s^{-1}$] & [$\rm km~s^{-1}$]  & \\
\hline
Bliss~1 & 0.44 & 0.72 & 1.16 & 3.43 & 4 \\
Bliss~1 & 0.42 & 0.88 & 1.11 & 3.07 & 6 \\
Kim ~3  & 0.25 & 0.93 & 3.12 & 20.69& 3 \\
\hline 
\end{tabular}
\tablefoot{The first row of the Table lists the four BLISS~1 stars of this study, while the second one lists additional two stars from the analysis by \citetalias{cerny_26}.}
\vspace{2pt}
\end{table}

Bliss~1: for this cluster we performed the Bayesian fit both on our sample (four stars) and also adding the two stars from the \citetalias{cerny_26} sample that are not in common with us (six stars). The mean $\rm V_{los}$ in the two cases is indistinguishable, hence we report only the result from the larger sample ${\rm V_{los}=123.0 \pm 0.6~km~s^{-1}}$, in excellent agreement with \citetalias{cerny_26} who found ${\rm V_{los}=122.6^{+1.0}_{-1.1}~km~s^{-1}}$. $\rm \sigma_{V_{los}}$ was partially resolved by \citetalias{cerny_26} who provides the 95th percentile of the PDF as an upper limit $\rm \sigma_{V_{los}}<5.3 {\rm ~km~s^{-1}}$. $\rm \sigma_{V_{los}}$ is only partially resolved also in our fits but the mode of the PDF is clearly different from zero and our $P_{95\%}$ upper limit is much more constraining than the one by \citetalias{cerny_26}, as we get $\rm \sigma_{V_{los}}<3.4 {\rm ~km~s^{-1}}$ from the smaller sample and $\rm \sigma_{V_{los}}<3.1 {\rm ~km~s^{-1}}$ from the larger one (see Tab.~\ref{tab:sigma}).

Kim~3: the fit was just a cross-check as all of our three stars are included in the \citetalias{cerny_26} sample and the uncertainties on individual $\rm V_{los}$ uncertainties are similar in the two studies, hence combining the two samples would be pointless. 
Our mean $\rm V_{los} = 164.5~^{+2.7}_{-3.0}~km~s^{-1}$ is in good agreement with \citetalias{cerny_26} $\rm V_{los} = 163.2^{+0.7}_{-0.8}~km~s^{-1}$, and, as expected, our upper limit on the velocity dispersion is much weaker than that derived by \citetalias{cerny_26}, $\rm \sigma_{V_{los}}<2.9 {\rm ~km~s^{-1}}$.

\section{Orbital parameters} \label{sect:op}

\begin{figure*}
\centering
\begin{minipage}{0.45\textwidth}
\centering        
\includegraphics[width=1.0\textwidth]{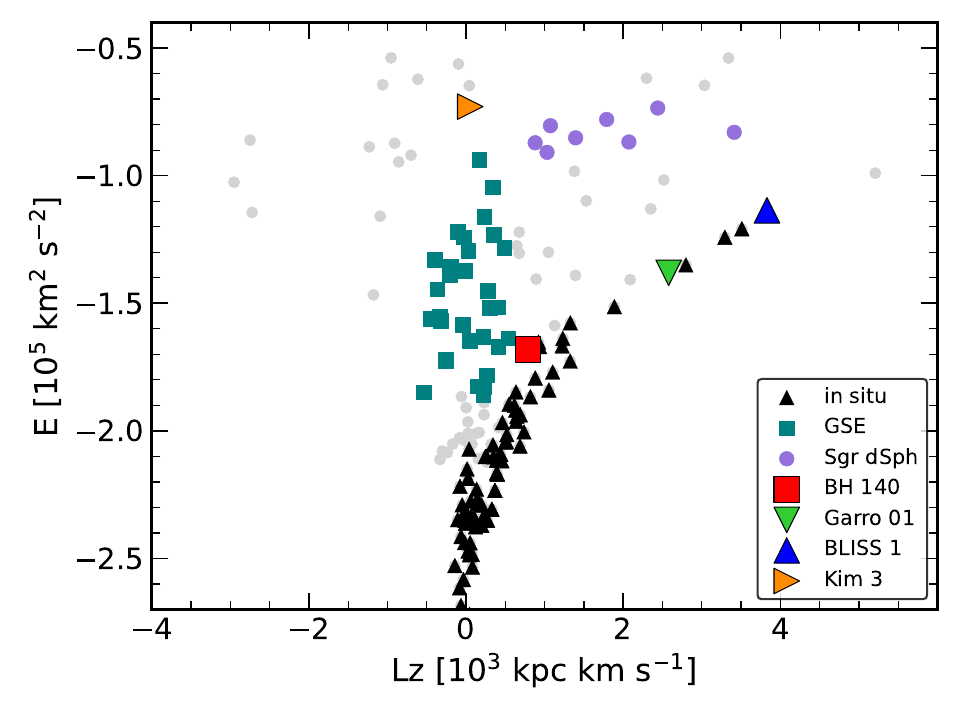}           
\end{minipage}
\begin{minipage}{0.45\textwidth}
\centering
\includegraphics[width=1.0\textwidth]{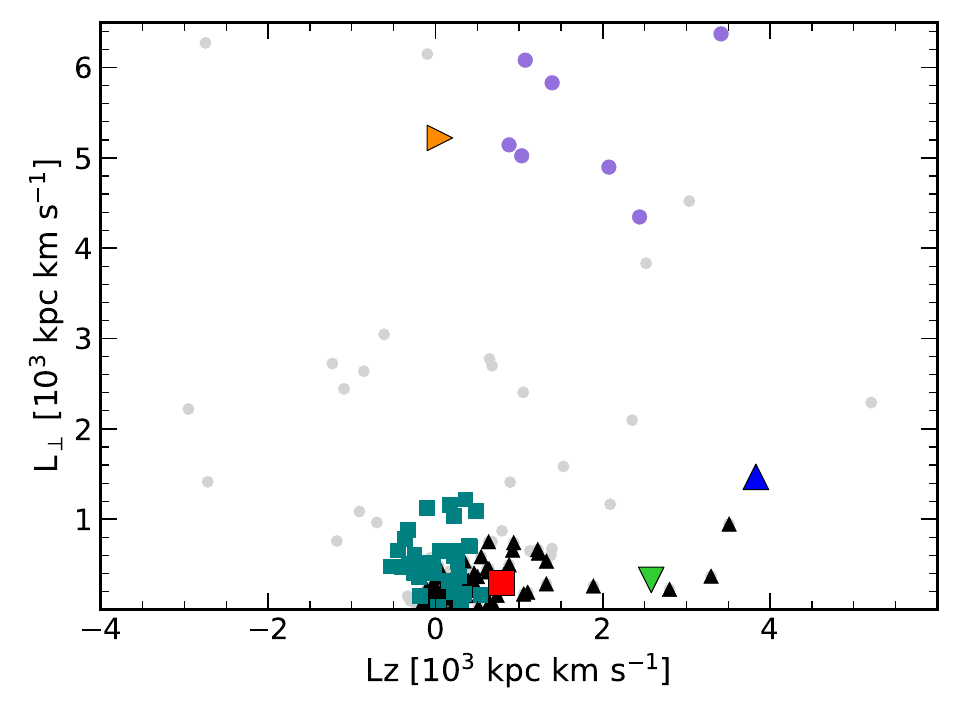}     
\end{minipage}  
\caption{Distribution in the integral-of-motions ($\rm E$, $\rm L_z$, $\rm L_{\perp}$) space of the stellar clusters analyzed in this paper, compared to all MW GCs from \citet[][eDR3 edition]{massari_19}.
The red square represents BH~140, the blue triangle BLISS~1, the inverted green triangle Garro~01, and the right-pointing orange triangle Kim~3. 
The black triangles represent the {\it in situ} GCs, the green squares the GCs associated with GSE, the purple circles the GCs associated with Sgr dSph, and the light-gray circles the GCs associated with other progenitors.
}
\label{fig:op}
\end{figure*}
\begin{table*}[h]
\caption{Orbital parameters for the observed clusters.}
\label{tab:op}
\centering
\renewcommand{\arraystretch}{1.5}
\normalsize
\setlength{\tabcolsep}{5.6pt}
\begin{tabular}{ccccccccc}
\hline\hline 
Cluster & Energy & $\rm L_{z}$ & $\rm L_{\perp}$ & $\rm r_{peri}$ & $\rm r_{apo}$ & $\rm Z_{max}$ & circ & ecc \\
 & [$\times10^{5} \; \rm km^{2} \; s^{-2}$]  & [$\times10^{3} \; \rm kpc \; km \; s^{-1}$] & [$\times10^{3} \; \rm kpc \; km \; s^{-1}$] & [kpc] & [kpc] & [kpc] &  \\
\hline
$\rm BH~140$  & $-1.68^{+0.05}_{-0.04}$ & $0.79^{+0.07}_{-0.06}$ & $0.29^{+0.01}_{-0.01}$ & $1.98^{+0.17}_{-0.16}$ & $10.15^{+1.01}_{-0.71}$ & $0.99^{+0.23}_{-0.16}$ & $0.52^{+0.01}_{-0.01}$ & 0.67 \\
$\rm Garro~1$ & $-1.38^{+0.06}_{-0.06}$ & $2.58^{+0.33}_{-0.45}$ & $0.33^{+0.02}_{-0.03}$ & $9.88^{+1.48}_{-2.20}$ & $13.28^{+1.54}_{-1.64}$ & $1.47^{+0.13}_{-0.22}$ & $0.96^{+0.01}_{-0.03}$ &  0.14 \\
$\rm Bliss~1$ & $-1.14^{+0.08}_{-0.06}$ & $3.83^{+0.77}_{-0.58}$ & $1.47^{+0.32}_{-0.24}$ & $15.72^{+4.93}_{-3.33}$ & $22.64^{+2.54}_{-1.94}$ & $8.14^{+0.58}_{-0.66}$ & $0.89^{+0.03}_{-0.04}$ & 0.18 \\
$\rm Kim~3$   & $-0.73^{+0.09}_{-0.10}$ & $0.05^{+0.05}_{-0.06}$ & $5.22^{+0.49}_{-0.59}$ & $12.56^{+0.70}_{-0.91}$ & $77.37^{+21.70}_{-18.57}$ & $65.78^{+6.73}_{-10.68}$ & $0.01^{+0.01}_{-0.01}$ & 0.72 \\
\hline 
\end{tabular}
\vspace{2pt}
\end{table*}

Information about the origin of Galactic GCs is encoded in their motion around the Galaxy, as the integrals of motion of their orbits may retain memory of their site of formation \citep{massari_19,callingham_22,garro_23,chen_24,deleo_26}.
To set up the initial conditions for integrating the orbits of our targets we coupled positions from \textit{Gaia} \citep{vasiliev_21} with proper motions obtained from the combination of \textit{Gaia} and \textit{HST} data from the MGCS survey \citep{libralato_26}, distances obtained from isochrone fitting \citep{baumgardt_21} and reported in the discovery papers \citep{kim_16,mau_19,garro_20}, and line-of-sight velocities from cluster spectra from this work. All the values are reported in Table~\ref{tab:info_cl}.
   
The cluster orbits were computed in a Galactocentric reference frame adopting the solar position and velocity from \citet{drimmel_18,gravity_18,bennett_19}.
Orbital integrations were carried out in the Milky Way potential described by \citet{mcmillan_17} using the \texttt{AGAMA}\footnote{The \texttt{AGAMA} library is written primarily in C++ to achieve maximum efficiency, and has a \texttt{PYTHON} interface offering greater flexibility for practical work.} package \citep{vasiliev_19}.
For each cluster we computed the integrals of motion, namely energy ($\rm E$), vertical angular momentum ($\rm L_z$), and perpendicular angular momentum ($\rm L_{\perp}$), the orbital pericenter ($\rm r_{peri}$) and apocenter ($\rm r_{apo}$) distances, the circularity (circ), and the maximum vertical excursion from the Galactic plane ($\rm Z_{max}$, \citealt{helmi_00,binney_08}).  Moreover, for each cluster we computed the eccentricity (ecc)\footnote{The eccentricity is a non-negative dimensionless parameter characterizing the non-circularity of a cluster’s three-dimensional trajectory, and it is defined as: $\rm ecc = \frac{r_{apo} - r_{peri}}{r_{apo} + r_{peri}}$.} of the orbit.

To account for observational uncertainties, we generated $100$ Monte Carlo realizations of the orbit of each cluster, sampling distance, proper motion, and systemic line-of-sight velocity assuming Gaussian errors. The orbital parameters were then derived from the resulting distributions, adopting the median values and their corresponding $\pm1\sigma$ intervals.

The position of the clusters in the $\rm E~vs~L_{z}$ and $\rm L_{\perp}~vs~L_{z}$ space is shown in Fig.~\ref{fig:op}, compared with the distribution of the {\it in situ} (triangles), {\it Gaia}-Sausage-Enceladus (GSE, \citealt{belokurov_18,helmi_18}, squares), Sagittarius dwarf spheroidal (Sgr dSph, \citealt{ibata_94}, dark-gray circles), and other progenitors (light-gray circles) GCs populations from \citet{massari_25_note}. 

The dynamics of BH~140 reveals a prograde orbit confined within the MW disk with a high eccentricity ($\rm ecc = 0.67$), with a maximum vertical height of $\rm Z_{max} =0.99~kpc$. 
Due to its highly eccentric orbit, BH~140 experiences significant radial excursions, ranging from the Galactic Bulge at $\rm r_{peri}=1.98~kpc$ to beyond the solar position at $\rm r_{apo}=10.15~kpc$. 
\citet{callingham_22} tentatively linked this system to the GSE merger event, even though they also suggested an alternative {\it in situ} disk origin. Also, \citet{petralia_26}, from their orbital analysis found that BH~140 is confined on a prograde orbital configuration, with a low vertical excursion ($\rm Z_{max}\lesssim1.52~kpc$) from the Galactic plane and a high eccentricity. 
Following the classification scheme proposed by \citet{massari_25_note}, we tentatively associate BH~140 to the {\it in situ} disk clusters, in agreement with the classification proposed by \citet{libralato_26}. However, this system is located in a region of the integrals of motion space very close to the GSE boundary, as its $\rm L_Z$ is slightly larger than the cutoff value proposed by \citet{massari_25_note}.  The chemical composition of this system will help to understand its origin. Indeed, it is well known that GSE GCs have lower $\rm [\alpha/Fe]$ compared to {\it in situ} GCs at fixed metallicity in the range $\rm -1.5\lesssim[Fe/H]\lesssim0.5$ \citep[e.g.,][]{horta_20,ceccarelli_24}.

The orbital properties of Garro~01 are fully consistent with an {\it in situ} origin \citep{massari_25_note}. We find it to reside on a prograde orbit with a low eccentricity ($\rm ecc = 0.14$) confined to the Galactic mid-plane ($\rm Z_{max}=1.47~kpc$) in the outer disk ($\rm r_{peri}=9.88~kpc$ and $\rm r_{apo}=13.28~kpc$), in agreement with parameters proposed by \citet{pace_23}. They classify Garro~01 as a likely {\it in situ} outer disk cluster. 

For BLISS~1, we derived a prograde orbit with a low eccentricity ($\rm ecc = 0.18$). This system is characterized by large vertical excursions ($\rm Z_{max}=8.14~kpc$) from the Galactic plane, with a $\rm r_{peri}$ beyond the solar orbit at 15.72~kpc and $\rm r_{apo}=22.64~kpc$. Such a high $\rm Z_{max}$ places BLISS~1 well outside the population of typical {\it in situ} disk clusters \citep{massari_25_note}. 
However, it should be noted that the MW disk is known to flare and warp in its outer parts \citep{poggio_25}. This makes a more generous $\rm Z_{max}$ cut reasonable to select outer {\it in situ} systems, such as those associated with the recently identified Aleph substructure, which represents the high-energy continuation of the Galactic disk \citep{deleo_26}. Therefore, while \citet{massari_25_note} classified BLISS~1 as a high-energy cluster lacking a clear progenitor association and \citet{callingham_22} proposed a dynamical link to the Helmi streams \citep{helmi_99}, a possible association with the heated outer disk (e.g., Aleph) remains a viable alternative. This dynamical ambiguity will be directly re-addressed in light of the chemical analysis performed in Sect.\ref{sect:chem_analysis_bliss1}.

Finally, Kim~3 occupies a high-energy orbital regime. Its orbit has a high eccentricity ($\rm ecc = 0.72$) and is characterized by a pericenter at $\rm r_{peri}=12.56~kpc$ and apocenter at $\rm r_{apo}=77.37~kpc$. 
The integrals of motion derived here are consistent with those of the Elqui stream \citep{ship_19}, and reported in \citet{massari_25_note}.

All the relevant orbital parameters derived in this work are summarized in Table~\ref{tab:op}. 
As a general caveat, it must be recalled that the available distance estimates for these cluster may suffer from some systematics, e.g., due to available photometry being too shallow, to the lack of Horizontal Branch stars and/or to the adoption of photometric metallicity significantly different from the spectroscopic values we derive here. 
In this sense, BLISS~1 represents the most extreme case. Indeed, a variation of $\rm \sim 0.7~dex$ in its metallicity determination (see Sect.~\ref{sect:chem_analysis_bliss1}) leads to a variation of $\sim2$~kpc in its distance. While systematic distance shifts primarily affect the exact values of the orbital energy and apocenter, the Monte Carlo realizations confirm that the overall dynamical classification remains robust against these variations.

\section{Chemical analysis} \label{sect:analysis}

\subsection{Linelist}
In this work we derived chemical abundances via spectral synthesis, adopting an appropriate linelist for each cluster. We compiled a list of suitable atomic lines over the entire spectral range covered by the X-shooter spectra. 
We selected transitions that were resolved, unblended, and devoid of significant telluric contamination and saturation effects at the instrumental resolution. 
While all spectra were corrected for telluric absorption using TAPAS synthetic models, deep telluric bands can still leave noisy residuals. Therefore, our line selection was performed avoiding transitions located in regions where significant telluric residuals or noise spikes were still present (e.g., the Fraunhofer B and A bands). 

\begin{figure*}
    \sidecaption
    \includegraphics[width=12cm]{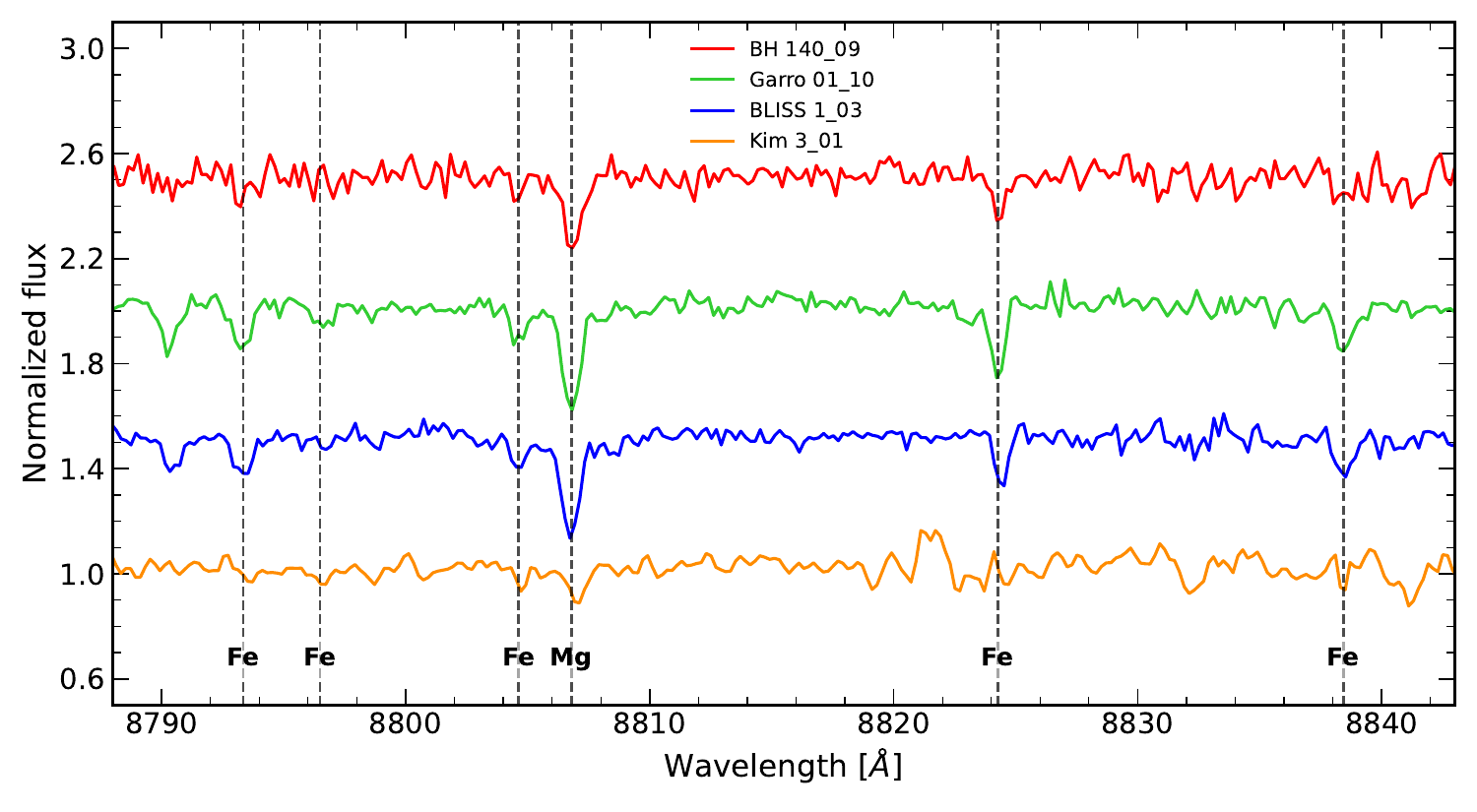}
     \caption{Portions of the X-shooter spectra for four stars of BH~140 (red solid line), BLISS~1 (blue solid line), Garro~01 (green solid line), and Kim~3 (orange solid line) around some lines of interest (Mg and Fe). The lines of interest for the chemical analysis are marked as dashed vertical lines.}
     \label{fig:spettri}
\end{figure*}

Our synthesis included a comprehensive set of atomic and molecular data from the Kurucz/Castelli database\footnote{\url{https://wwwuser.oats.inaf.it/castelli/linelists.html}}, integrated with updated $\rm \log gf$ values from recent laboratory studies. 
We note that all the rest-frame wavelengths adopted throughout this work are given in air.
As an example, Fig.~\ref{fig:spettri} shows a portion of X-shooter spectra for four stars belonging to BH~140, BLISS~1, Garro~01, and Kim~3, in which we have marked some lines of interest  that were used during the analysis.
The final linelist included lines of Na, Mg, Al, Si, Ca, Fe, and Ba.
In the following paragraphs we describe the specific lines adopted for each chemical species.

We derived Fe abundances for all the targeted clusters, employing a number of Fe lines ranging from five to 60, depending on the spectral quality of each individual star.
Regarding the $\alpha$-elements, we successfully obtained Mg abundances for all the analyzed systems, with the exception of Kim~3, using the transition at $8806.7~\AA$. To properly account for departures from local thermal equilibrium (LTE), we applied the appropriate non-LTE (NLTE) corrections by interpolating the grids provided by \citet{bergemann_17} through the MPIA database\footnote{\url{https://nlte.mpia.de/gui-siuAC_secE.php}}. Furthermore, we measured Ca for BLISS~1 and Garro~01 stars using up to seven lines distributed across the X-shooter spectral range, while Si abundances were derived exclusively for BLISS~1 using up to five available transitions.
Among the light elements, Na was derived only for BLISS~1 using the doublet at $5889.95$ and $5895.92~\AA$. For the remaining clusters, reliable Na abundances could not be determined due to heavy blending with telluric lines or insufficient SNR in this spectral region. Similar to Mg, we corrected the Na abundances for NLTE effects using the grids from \citet{lind_11} via the INSPECT database\footnote{\url{https://inspect-stars.com/}}. Aluminum, instead, was measured only for the stars in Garro~01 relying on the doublet located at $7835-7836~\AA$.
Finally, we derived abundances for the neutron-capture element Ba in both BLISS~1 and Garro~01, using the absorption line at $6141.7~\AA$.

All abundance ratios reported in this study are scaled to the solar chemical composition by \citet{lodders_10} and \citet{caffau_11}.

\subsection{Abundance analysis}
Chemical abundances were derived using the \texttt{SALVADOR} (Alvarez Garay et al., in prep.) code, which determines the best-fit abundance through a $\chi^2$ minimization between the observed line profile and a grid of synthetic spectra. These grids were generated by varying the abundance of the considered element and using the appropriate stellar atmospheric parameters for each star, as derived in Sect.~\ref{sect:param_atm}. The underlying synthetic spectra were computed with the \texttt{SYNTHE} suite, based on 1D plane-parallel LTE model atmospheres produced with the \texttt{ATLAS9} code \citep{sbordone_04,kurucz_05}. To ensure high-quality spectral synthesis, we employed the latest opacity distribution functions provided by the KOALA database\footnote{\url{https://sites.google.com/view/koala-database/}} \citep{mucciarelli_26}.

For elements with multiple measured features, the final stellar abundance was computed as the simple mean of the abundances derived from the individual spectral lines. For Mg and Ba, the final abundance was the one derived from the single spectral feature used.

Uncertainties associated with the chemical abundances were calculated as the sum in quadrature of the error related to measurement process (systematic) and the errors associated to the atmospheric parameters (internal), following the methodology outlined in \citet{alvarez_22}.
Systematic errors originating from the atmospheric parameters were quantified by perturbing $\rm T_{eff}$, $\rm \log g$, and $\rm v_t$ individually by their respective $1\sigma$ uncertainties and recalculating the abundances. The internal uncertainty, reflecting the precision of the measurement process, was derived from the line-to-line scatter, normalized by the square root of the number of lines used. For elements where the abundance relied on a single transition (e.g., Mg), we performed Monte Carlo simulations to obtain a robust error estimate. We generated 200 synthetic realizations of the specific line, adding Poissonian noise to the synthetic spectrum consistent with the observed SNR. Each simulated spectrum was analyzed with the same procedure used for observed spectra; the resulting standard deviation of these 200 iterations was then adopted as the internal measurement error.

It is important to note that the observational limitations imposed by a low SNR are quantitatively accounted for in our error budget. Because the internal uncertainties are derived either from the line-to-line scatter or through Monte Carlo simulations that explicitly replicate the observed spectral noise, a poorer data quality directly propagates into larger, more conservative errors for the derived abundances.
Based on spectral synthesis simulations, a minimum SNR of $\sim15-20$ per resolution element is required to reliably identify and fit the strongest transitions with a precision better than $\rm \sim0.15~dex$. Below this threshold, blending with noise spikes prevents accurate abundance determinations, as was the case for the two faintest targets in Kim~3.

\section{ Abundance results} \label{sect:res_chem}

In this section, we present the chemical abundances derived for the four stellar systems under analysis.
The resulting mean abundances and their dispersions are summarized in the following subsections and the average value of each abundance ratio (and the corresponding uncertainty) for all the stars observed in each cluster is listed in Table~\ref{tab:3}.
The mean values and intrinsic standard deviations for both the metallicity ($\rm \langle [Fe/H]\rangle$, $\rm \sigma_{[Fe/H]}$) and the abundance ratios ($\rm \langle [X/Fe]\rangle$, $\rm \sigma_{[X/Fe]}$), along with their associated uncertainties, were computed by means of the Maximum Likelihood technique described in \citet{muccia12}.
This approach properly weights each stellar measurement by its individual total uncertainty, providing a robust estimate of the systemic cluster abundance, the intrinsic abundance spread, and their respective propagated errors.

The mean value of each element, for all the clusters, is reported in Table~\ref{tab:3}.

\begin{table*}[h]
\caption{Chemical abundances of the target stars.}
\label{tab:3}
\centering
\renewcommand{\arraystretch}{1.5}
\normalsize
\setlength{\tabcolsep}{5.6pt}
\begin{tabular}{cccccccc}
\hline\hline 
ID & [Fe/H] & [Na/Fe] & [Mg/Fe] & [Al/Fe] & [Si/Fe] & [Ca/Fe] & [Ba/Fe] \\
 & 7.52 & 6.29 & 7.54 & 6.46 & 7.53 & 6.31 & 2.18 \\
\hline

$\rm BH~140\_04$ & $-0.71\pm0.26$ & ... & $0.00\pm0.15$ & ... & ... & ... & ... \\		       
$\rm BH~140\_06$ & $-0.88\pm0.24$ & ... & $0.30\pm0.11$ & ... & ... & ... & ... \\
$\rm BH~140\_07$ & $-0.80\pm0.26$ & ... & $0.15\pm0.17$ & ... & ... & ... & ... \\
$\rm BH~140\_09$ & $-0.85\pm0.24$ & ... & $0.25\pm0.16$ & ... & ... & ... & ... \\
Mean value & $-0.81\pm0.12$ & ... & $0.20\pm0.07$ & ... & ... & ... & ... \\ 
\hline
$\rm Garro~01\_02$ & $-0.73\pm0.11$ & ... & $0.45\pm0.09$ & $ 0.63\pm 0.07$ & ... & $0.29\pm0.11$ & $ 0.17\pm 0.10$ \\
$\rm Garro~01\_09$ & $-0.75\pm0.17$ & ... & $0.52\pm0.08$ & $     ...     $ & ... & $0.45\pm0.13$ & $ 0.41\pm 0.12$ \\
Mean value & $-0.74\pm0.09$ & ... & $0.48\pm0.06$ & ... & ... & $0.35\pm0.08$ & $0.27\pm0.09$ \\
\hline
$\rm BLISS~1\_01$ & $-0.66\pm0.11$ & $0.31\pm0.11$ & $0.25\pm0.07$ & ... & $ 0.23\pm0.15$ & $0.16\pm0.13$ & $0.22\pm0.08$ \\
$\rm BLISS~1\_02$ & $-0.64\pm0.11$ & $0.32\pm0.09$ & $0.36\pm0.08$ & ... & $ 0.14\pm0.17$ & $0.28\pm0.09$ & $0.21\pm0.09$ \\
$\rm BLISS~1\_03$ & $-0.65\pm0.13$ & $0.26\pm0.08$ & $0.23\pm0.08$ & ... & $ 0.27\pm0.13$ & $0.33\pm0.12$ & $0.29\pm0.09$ \\
$\rm BLISS~1\_04$ & $-0.67\pm0.14$ & $0.29\pm0.08$ & $0.39\pm0.07$ & ... & $ ...        $ & $0.31\pm0.12$ & $0.42\pm0.12$ \\
Mean value & $-0.65\pm0.06$ & $0.29\pm0.04$ & $0.31\pm0.04$ & ... & $0.22\pm0.08$ & $0.28\pm0.05$ & $0.27\pm0.04$ \\
\hline
$\rm Kim~3\_01$ & $-1.78\pm0.30$ & ... & ... & ... & ... & ... & ... \\

\hline 
\end{tabular}
\tablefoot{The adopted solar abundances for the measured chemical elements are from \citet{lodders_10} and \citet{caffau_11} and they are reported in the header of each element abundance column. For each cluster, the mean elemental abundances are reported immediately below the individual stellar measurements.}
\end{table*}

\subsection{BH~140} \label{sect:res_chem_bh140}
\begin{figure}[h]
\centering
\includegraphics[width=9 cm]{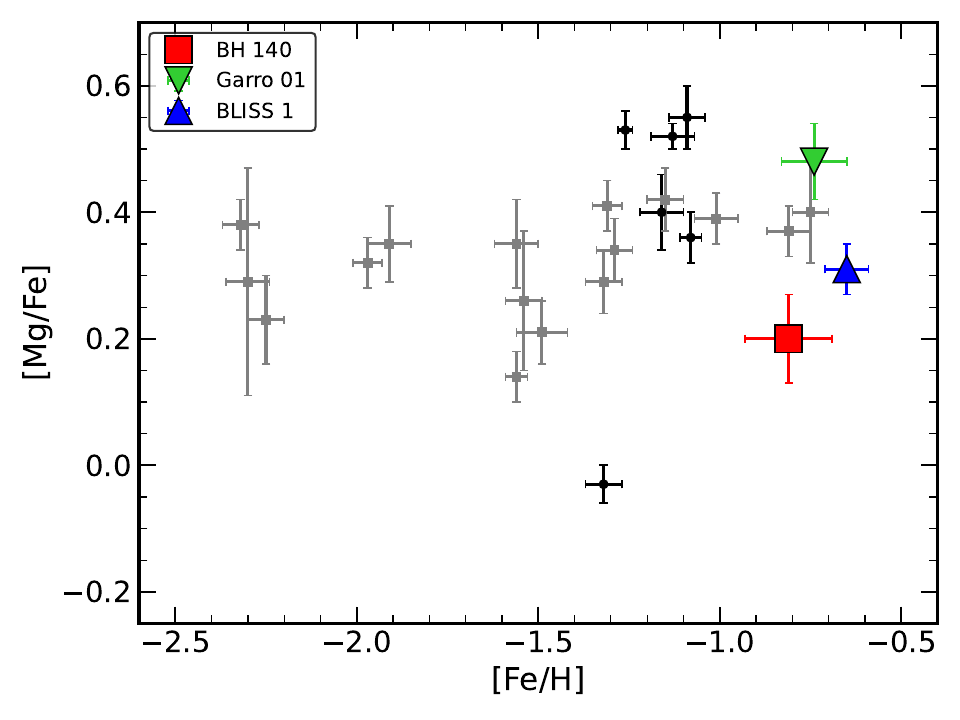}
\caption{Run of the mean [Mg/Fe] as a function of mean [Fe/H] for BH~140 (red square), BLISS~1 (blue triangle), and Garro~01 (green triangle). As a comparison, the same trend is plotted for 22 GCs taken from the works by \citet[][black squares]{ceccarelli_24} and \citet[][gray circles]{carretta_26}.
For each cluster, the error bar indicates the $1\sigma$ spread.} \label{fig:mg_fe}
\end{figure}

For BH~140 no chemical information is available in the literature.
Therefore, its chemical characterization is essential to determine its nature and to place it within the context of the Galactic cluster population.
We derived chemical abundances for four stars of BH~140. 
We found $\rm \langle [Fe/H]\rangle = -0.81\pm0.12$~ and  $\rm ~\sigma_{[Fe/H]}=0.00\pm0.13~dex$, consistent with no metallicity spread. 
Regarding the $\rm \alpha$-elements, we derived Mg abundances for all the targets, finding $\rm \langle [Mg/Fe]\rangle = 0.20\pm0.07~$ and $\rm \sigma_{[Mg/Fe]}=0.00\pm0.15~dex)$. 
In Fig.~\ref{fig:mg_fe} we show the behavior of the mean [Mg/Fe] abundance ratio as a function of [Fe/H]. There is an indication that BH~140 displays a lower Mg abundance compared to other Galactic GCs at similar metallicity ($\rm \langle [Mg/Fe]\rangle \sim 0.45$; \citealt{ceccarelli_24,carretta_26}).
We scaled the abundance ratios by \citet{ceccarelli_24} and \citet{carretta_26} to the same solar reference that we employed in this work \citep{lodders_10,caffau_11}. 
It is worth noting that any residual offset between our results and those of \citet{carretta_26} is likely driven by the different adopted effective temperature scales, while our measurements and those by \citet{ceccarelli_24} adopt the same temperature scale.

Among the observed individual targets, the star BH~140\_04 exhibits a nominally lower Mg abundance ($\rm [Mg/Fe] = 0.00\pm0.15~dex$) compared to other stars. While this star is the coolest in our sample, its stellar parameters and [Fe/H] remain fully consistent with the other members within the uncertainties.
This apparent Mg depletion is not statistically significant, because due to the measurement errors, its abundance is compatible with the rest of the sample within $1.6\sigma$.
Nevertheless, to ensure that the low Mg content of the cluster is not driven by this single measurement, we recalculated the mean excluding BH~140\_04. The resulting value ($\rm [Mg/Fe] = 0.25\pm0.08~dex$) is marginally higher, but it remains somewhat lower than the typical $\alpha$-enhancement observed in classical GCs at similar metallicity \citep[e.g.,][]{ceccarelli_24,carretta_26}. 

The low dispersion derived for $\rm [Mg/Fe]$ points towards a chemically homogeneous population constituting BH~140. However, since the available spectra did not allow us to measure other key light elements (such as Na or O), we cannot currently rule out the presence of the internal chemical spreads typically associated with multiple populations in classic GCs.

\subsection{Garro~01} \label{sect:chem_garro1}

In this work we analyzed two stars belonging to Garro~01. We measured $\rm \langle [Fe/H]\rangle = -0.74\pm0.09$~ and  $\rm \sigma_{[Fe/H]}=0.00\pm0.09~dex$. 
This result is in excellent agreement with the value found by \citet{garro_20}, who estimated a metallicity of $\rm [Fe/H]=-0.7\pm0.2~dex$ from isochrone fitting and photometric comparisons with the GC 47~Tuc. 
Our spectroscopic metallicity is lower than the value reported by \citet{pace_23}, who found $\rm \langle [Fe/H]\rangle = -0.30\pm0.03$~ and $\rm \sigma_{[Fe/H]}<0.14~dex$ using high-resolution spectra obtained with Michigan/Magellan Fiber System \citep{mateo_12} at the Magellan/Clay Telescope at Las Campanas Observatory (Chile). However, from their Figs.~1 and 4 is possible to observe that their Fe distribution spans a wide range of [Fe/H], from $\sim -0.7$ up to $\rm \sim0~dex$. 
A cross-match with the sample analyzed by \citet{pace_23} reveals one star in common (Garro~01\_09). Their derived metallicity for this target ($\rm [Fe/H]=-0.45\pm0.14~dex$) is consistent with our measurement within $1.35\sigma$. 

Garro~01 exhibits a significant enhancement in $\rm \alpha$-elements, with
$\rm \langle [Mg/Fe]\rangle = 0.48\pm0.06$~ and  $\rm \sigma_{[Mg/Fe]}=0.00\pm0.06~dex$.
Our Mg abundance for Garro~01\_09 differs from the value reported by \citet{pace_23} ($\rm [Mg/Fe] = 0.25\pm0.09~dex$) by $\sim2.2\sigma$. This offset is likely driven by the different analysis techniques, the adopted lines and the treatment of NLTE effects.
The observed lack of a Mg spread does not unequivocally discriminate between an open and a globular cluster, as a null spread is theoretically expected for both classes in this mass and metallicity regime (see \citealt{dal_ponte_25} for open clusters and \citealt{shetrone_96} for multiple populations in GCs\footnote{The onset of the Mg-Al anticorrelation in GCs starts at $\rm [Fe/H]<-1~dex$ \citep{shetrone_96} and reach its maximum extension at $\rm [Fe/H]\sim-1.5~dex$ \citep{pancino_17}.}). However, its overall metallicity is far more typical of a genuine GC, but a definitive classification will require the measurement of key elements such as C, N, O, and Na that are involved in the phenomenon of multiple populations in GCs \citep[e.g.,][]{carretta_09}.

Regarding Ca, we measured $\rm \langle [Ca/Fe]\rangle = 0.35\pm0.08$ and  $\rm \sigma_{[Ca/Fe]}=0.00\pm0.11~dex$. 
We were able to measure Al abundance for one star, Garro~01\_02, that has $\rm [Al/Fe]=0.63\pm0.07$. 
Finally, we were able to measure also Ba abundances. In this case we obtained 
$\rm \langle [Ba/Fe]\rangle = 0.27\pm0.09$ and  $\rm \sigma_{[Ba/Fe]}=0.04\pm0.18~dex$.
The low intrinsic dispersion in all the measured abundance ratios indicates that Garro~01 hosts a chemically homogeneous stellar population.
All the abundance patterns for Garro~01 are shown in Fig.\ref{fig:abu_pattern}, where they are compared to those of the other stellar systems.

\subsection{BLISS~1} \label{sect:chem_analysis_bliss1}
BLISS~1 is an ultra-faint stellar system that \citet{mau_19} classified as a halo star cluster rather than a UFD galaxy.
However, a spectroscopic study of this system is necessary to confirm this
classification through systemic velocity, metallicity dispersion measurements, and elemental abundance determination.

We analyzed four likely member stars of BLISS~1. The iron content was found to be $\rm \langle [Fe/H]\rangle = -0.65\pm0.06$ and  $\rm \sigma_{[Fe/H]}=0.00\pm0.06~dex$. 
This value is significantly higher than the estimate provided by \citet{mau_19}, and is independently supported by the recent results of \citetalias{cerny_26}. From the analysis of three member stars, they found a metallicity of $\rm [Fe/H]=-0.83\pm0.14~dex$. Although this value is slightly lower than our determination, the two measurements are fully compatible within $1.2~\sigma$.
Furthermore, \citetalias{cerny_26} derived a significantly younger age for BLISS~1 compared to the initial estimate by \citet{mau_19}, reporting an age of $\rm 7~Gyr$.

\begin{figure}[h]
\centering
\includegraphics[width=9 cm]{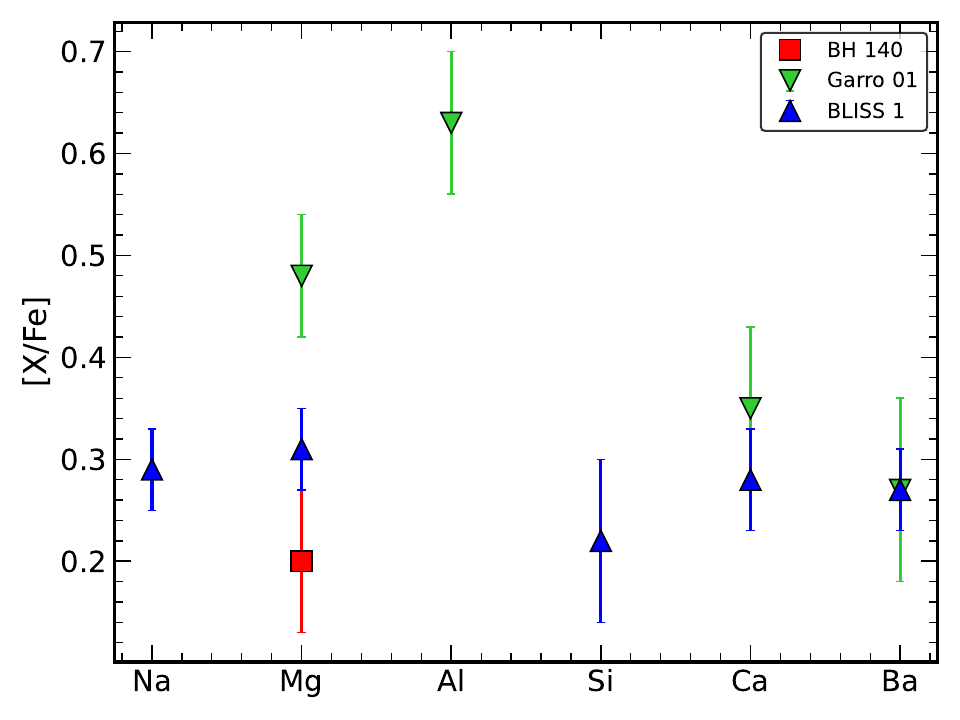}
\caption{Mean chemical abundance patterns for BH~140 (red square), BLISS~1 (blue triangles), and Garro~01 (green triangles). The plot displays the average [X/Fe] ratios for the light (Na, Al), $\alpha$- (Mg, Si, Ca), and neutron-capture (Ba) elements derived in this work. Error bars represent the 1$\sigma$ uncertainties associated with the measurements.} \label{fig:abu_pattern}
\end{figure}

BLISS~1 exhibits an enhancement in $\rm \alpha$-elements (for Mg see Fig.\ref{fig:mg_fe}). We were able to derive abundances for Mg, Si, and Ca. We measured $\rm \langle [Mg/Fe]\rangle = 0.31\pm0.04$ and  $\rm \sigma_{[Mg/Fe]}=0.00\pm0.11~dex$, $\rm \langle [Si/Fe]\rangle = 0.22\pm0.08$ and  $\rm \sigma_{[Si/Fe]}=0.00\pm0.09~dex$ (based on three stars), and $\rm \langle [Ca/Fe]\rangle = 0.28\pm0.05$ and  $\rm \sigma_{[Ca/Fe]}=0.00\pm0.06~dex$. All their values are reported in Fig~\ref{fig:abu_pattern}.
These values are compatible with those usually measured in Galactic GCs \citep[e.g.,][]{carretta_26}. 

BLISS~1 appears to be a metal-rich, $\alpha$-enhanced and possibly young system. 
Thus, it is instructive to compare it with other GCs in a similar metallicity regime.
On the one hand, young systems such as Palomar~12 (age $\rm \sim9~Gyr$; \citealt{musella_18}) and Terzan~7 (age $\rm \sim6~Gyr$; \citealt{taut_04}) have comparable metallicities ($\rm [Fe/H]\sim-0.8$ and $\rm -0.6~dex$, respectively; \citealt{cohen_04,sbordone_05}) but do not show $\alpha$-enhancement. Indeed, they have originated within the Sgr dSph system, which was characterized by a slow star formation rate \citep[e.g.,][]{sbordone_05}, and were subsequently stripped by the MW \citep[][and references therein]{bellazzini_20}.  
On the other hand, classical, old Galactic GCs at similar metallicity, such as 47~Tuc ($\rm [Fe/H]\sim-0.75~dex$), are $\alpha$-enhanced \citep{pancino_17,carretta_26}.
Even when accounting for potential systematics among different studies (such as the adopted temperature scales), the chemical signature of BLISS~1 closely resembles that of the old GCs like 47~Tuc rather than the young Palomar~12 and Terzan~7, possibly suggesting an {\it in situ} origin (see Sect.~\ref{sect:discussion_bliss1} for a discussion). 

We also derived abundances for Na, finding $\rm \langle [Na/Fe]\rangle = 0.29\pm0.04$ and  $\rm \sigma_{[Na/Fe]}=0.00\pm0.04~dex$. 
Also in this case a comparison with the Na abundances of Palomar~12 and Terzan~7 shows a clear difference. Indeed, these systems are characterized by sub-solar [Na/Fe], with Palomar~12 and Terzan~7 having $\rm [Na/Fe]\sim-0.50$ and $\rm\sim-0.15~dex$, respectively \citep{cohen_04,taut_04}.  
On the other hand, 47~Tuc has $\rm \langle[Na/Fe]\rangle \sim 0.35$ \citep{cordero_14,johnson_15}, a value compatible with that of BLISS~1. 

Similarly to $\alpha$ and Na, the neutron-capture element Ba shows an enhancement, with $\rm \langle [Ba/Fe]\rangle = 0.27\pm0.04$ and  $\rm \sigma_{[Ba/Fe]}=0.00\pm0.06~dex$.
All the abundance ratios of BLISS~1 are illustrated in Fig.\ref{fig:abu_pattern}.

The low dispersion in all derived abundance ratios points towards a chemically homogeneous stellar population constituting BLISS~1. The combined evidence thus points towards an {\it in situ} system that shows no evidence for light-elements internal spreads.

\subsection{Kim~3}
While recent spectroscopic studies have provided $\rm V_{los}$ measurements for Kim~3 \citep{cerny_26}, its extreme faintness has so far prevented any direct determination of its chemical composition. Obtaining its metallicity, alongside $\rm V_{los}$ measurements, is therefore crucial to establish its physical nature, accurately reconstruct its orbital history, and understand its survival within the Galactic tidal field.
Among the three stars belonging to Kim~3 presented here, we were able to obtain a direct measure of the iron content only for one star, namely Kim~3\_01\footnote{The SNR of this star is around 10, and we were able to use just five Fe lines to derive its metallicity.}. This star has $\rm [Fe/H] = -1.78\pm0.30$ dex. This value is compatible with the metallicity found from isochrone fitting by \citet{kim_16}, who found $\rm [Fe/H] = -1.6^{+0.45}_{-0.30}~dex$. 
Unfortunately, the SNR of the other two stars in our sample is too low ($< 10$) to reliably derive individual Fe abundances from metallic lines. Furthermore, although the near-infrared Calcium Triplet (CaT) lines are clearly visible in all the acquired spectra, we could not exploit empirical CaT--[Fe/H] relations to estimate the metallicity of these two targets. This limitation arises because the stars are located near the main-sequence turn-off, a region where standard CaT--[Fe/H] calibrations are currently not defined.

\section{Discussion and conclusions} \label{sect:disc}
In this work, we presented a medium-resolution spectroscopic analysis of stars of the stellar systems BH~140, Garro~01, BLISS~1, and Kim~3, derived from VLT/X-shooter observations. By combining precise $\rm V_{los}$ and chemical abundances, we confirmed the membership of the analyzed targets and provided the first chemical characterization of BH~140, Garro~01, and BLISS~1 as well as the first spectroscopic metallicity measurement for Kim~3.
Our analysis also provides an independent validation of the $\rm V_{los}$ and metallicities recently reported in the literature \citep[][\citetalias{cerny_26}]{pace_23, petralia_26}.

Metallicity and detailed chemical abundance patterns are fundamental properties that serve a dual purpose for poorly studied stellar systems, like those analyzed in this work. First, they are crucial for a robust physical classification, allowing to distinguish a faint star cluster from an UFD galaxy, or a globular from an open cluster \citep[see, e.g.,][\citetalias{cerny_26}]{pace_23,petralia_26}.
Second, they act as diagnostic tools to trace their birth environments, allowing to determine whether they formed {\it in situ} (e.g., within the Galactic disk or bulge) or were accreted from external galactic building blocks \citep[see, e.g.,][and references therein]{massari_19,ceccarelli_24}.

In the following, we discuss the properties of each analyzed system in light of our new spectroscopic findings and the existing literature.

\subsection{BH~140: a metal-rich {\it in situ} or accreted cluster?}
 
The relatively high metal content of BH~140, coupled with the lack of metallicity spread, strongly supports the idea that BH~140 is a genuine star cluster. Its structural parameters lie within the distribution of MW GCs, in agreement with the GC classification by \citet{cantat_18}. The low value of $\rm Z_{max}$ suggests an {\it in situ} origin, but the integrals of motion are at the boundaries of the region enclosing the GCs attributed to the GSE remnant and the high orbital eccentricity may support an accreted origin \citep[see also][]{petralia_26}. 
BH~140 also shows a mildly lower degree of $\alpha$ enhancement with respect to other GCs. This may support an accreted origin, but the abundance of other elements sensitive to a cluster origin \citep[e.g., Zn, Eu;][]{minelli_21,monty_24} are required to obtain a firmer conclusion. In passing, we note that if BH~140 were associated with the GSE merger, it would represent the most metal-rich cluster linked to this event. While currently known
GSE GCs are more metal-poor than BH~140, its relatively high metallicity does not exclude this association. Indeed, the GSE field star population has been observed to extend up to metallicities of $\rm [Fe/H] \sim -0.5~dex$ or potentially even higher \citep{myeong_19,feuillet_21,gonzalez_25,Kushniruk_26}. Finally, in case of an accreted origin from GSE and according to the age-metallicity relation derived by \citet{aguado_agelet_25}, BH~140 should have an age of $\rm \sim10~Gyr$ or lower.

\subsection{Garro~01: open or globular cluster?}
In their work, \citet{garro_20} classified Garro~01 as a low-luminosity, metal-rich GC.
However, the true nature of this system has been recently debated by \citet{pace_23}, who found $\rm [Fe/H]= -0.30~dex$ and an age of $\rm \sim 4~Gyr$, suggesting that Garro~01 could be an open cluster rather than a GC. 

Our spectroscopic analysis yields a $\rm \langle[Fe/H]\rangle=-0.74\pm0.09~$dex, which is in excellent agreement with the original photometric estimate by \citet{garro_20} and lower than the value found by \citet{pace_23}. More importantly, our chemical characterization reveals abundance patterns that are highly unusual for standard Galactic open clusters \citep{spina_21,carbajo_24,dal_ponte_25}. 
Specifically, we detected a remarkable enhancement in Al in one star ($\rm [Al/Fe]=+0.63~ dex$), alongside with enhanced abundances of $\alpha$-elements (Mg and Ca). 
As recently demonstrated by large-scale, high-precision spectroscopic surveys of the Galactic disk \citep[e.g.,][]{spina_21,carbajo_24,dal_ponte_25}, typical open clusters do not exhibit such high Al and $\alpha$ abundances. 
Instead, these chemical signatures are more likely observed in GCs \citep{carretta_13,carretta_26}.
Finally, the Ba abundances measured point toward an enrichment of this system by low-mass asymptotic giant branch stars \citep{gallino_98,travaglio_04,karakas_14}. 

Based on our chemo-dynamical analysis, we tentatively classify Garro~01 as an {\it in situ} GC rather than an open cluster. However, its definitive classification remains limited by conflicting age estimates in the literature: while \citet{garro_20} qualitatively suggested an old age of $\sim11$~Gyr based on a visual comparison with 47 Tuc, \citet{pace_23} explicitly derived a much younger age of $\sim 4$~Gyr. Resolving this discrepancy requires a robust age determination via deep CMD fitting down to the main sequence. Therefore, combining future deep photometry from the MGCS with chemo-dynamical data from a larger stellar sample will be the crucial step to unambiguously reveal the true nature of this system.

\subsection{BLISS~1: a young $\alpha$-enhanced cluster?} \label{sect:discussion_bliss1}
Discovered by \citet{mau_19}, BLISS~1 was initially classified as an ultra-faint, low-mass star cluster residing in the outer halo. Based on isochrone fitting, \citet{mau_19} estimated a metal-poor composition ($\rm [Fe/H]=-1.4~$dex) and an intermediate age of 9.2~Gyr (see also \citetalias{cerny_26}). 
However, our direct spectroscopic measurements revise this picture, revealing a much higher metallicity ($\rm [Fe/H]=-0.65~dex$; in good agreement with \citetalias{cerny_26}) and negligible metallicity spread. 
The confirmation of such a high metallicity in a system with an extremely low present-day luminosity places BLISS~1 completely off the mass-metallicity relation expected for dwarf galaxies \citep{kirby_13,simon_19,fu_23,geha26_I}, supporting its classification as a faint star cluster (see Fig.~\ref{fig:rh_mv}). 
Moreover, the newly derived metallicity certainly has implications on the age and distance derived by \citet{mau_19} assuming $\rm [Fe/H]=-1.4~$. In particular, based on a preliminary isochrone fitting of the MGCS photometry that adopts our newly derived metallicity, BLISS~1 appears to be somewhat younger than the estimate reported by \citet[][Dodd et al. in prep.]{mau_19}. 
A younger age would have important implications for the origin of BLISS~1, tentatively favoring an {\it in situ} origin (e.g., as an old open cluster formed in the Galactic disk) over a classic accreted halo GC scenario. However, a quantitative derivation of its age and a comprehensive discussion of its evolutionary history are deferred to an upcoming work (Dodd et al., in prep.).
Further evidence against an UFD galaxy classification comes from Ba abundances. Indeed, UFD galaxies display systematically lower abundances of the neutron-capture elements Sr and Ba compared to GC counterparts (e.g., $\rm [Sr(Ba)/Fe]<-1~dex $, \citealt{simon_10,frebel_10,koch_13,frebel_15,ji_19,webber_23}).

BLISS~1 exhibits enhancements in $\alpha$-elements and Na perfectly mirroring the abundance patterns of classical GCs. Notably, the system lacks any measurable internal spread in $\alpha$-elements and displays a striking uniformity in Na. This homogeneity strongly suggests that BLISS~1 does not host the multiple populations usually observed in massive GCs.
This absence is not unexpected, since it was proposed that multiple populations are present in intermediate-age GCs with masses higher than $\rm \sim5\times10^4~M_{\odot}$ \citep{li_19}, a threshold far above the extremely low present-day stellar mass of $\rm \sim140~M_{\odot}$ estimated for BLISS~1 \citep{mau_19}.

While the kinematics of BLISS~1 formally place it within the high-energy group of clusters \citep{massari_25_note}, its true origin requires an updated interpretation, based on the findings presented in this work. 
Indeed, its large vertical excursion from the Galactic plane ($\rm Z_{max}=8.14~kpc$) was used by \citep{massari_25_note} as a criterion to exclude it from the sample of {\it in situ} GCs. However, consistently with the chemical evidence we find here, an {\it in situ} origin becomes now viable if we interpret BLISS~1 as linked to the flare and warp of the outer disk \citep{poggio_25}.
In this context, BLISS~1 could be dynamically associated with Aleph, a high-energy substructure recently interpreted as a direct continuation of the Galactic disk \citep[e.g.,][]{deleo_26}, that was not included by \citet{massari_25} in their list of progenitors. As demonstrated by \citep{horta_23}, Aleph exhibits chemical abundance patterns typical of {\it in situ} disk populations, firmly distinguishing it from accreted dwarf galaxy debris. 

In summary, our combined chemo-dynamical evidence demonstrates that BLISS~1 is a genuine, chemically homogeneous star cluster rather than a UFD galaxy. Its metal-rich, $\alpha$-enhanced composition and the lack of multiple populations strongly support an {\it in situ} formation scenario. Furthermore, its high-energy orbit and substantial vertical excursion make it a prime candidate for an {\it in situ} cluster dynamically heated by the flaring and warping of the outer Galactic disk, naturally associating it with the Aleph substructure

\subsection{Kim~3: a faint metal-poor star cluster?}
The direct spectroscopic metallicity obtained for Kim~3 ($\rm [Fe/H]=-1.78\pm0.30~dex$), even though based on only one star, is in good agreement with the original photometric estimate by \citet{kim_16} and places the system outside the typical metallicity range of UFD galaxies (\citetalias{cerny_26}). 
Interestingly, if the intermediate age estimate by \citet[][$\sim 9.5$~Gyr; see also \citetalias{cerny_26}]{kim_16} will be confirmed, to our knowledge it would be the first case of such a metal-poor cluster significantly younger than $\sim 12$~Gyr \citep[see, e.g.,][]{dotter10,vande13,massari_26}.

Based on our updated kinematics, the integrals of motion of Kim~3 fall within the range covered by the stars of the Elqui stream, rather than aligning with those of the Sgr dSph \citep[see][]{massari_25_note}. However, future studies targeting a larger stellar sample will be crucial to fully unravel the history of this system.
\newline

In summary, the present analysis follows recent literature studies \citep[][\citetalias{cerny_26}]{pace_23, massari_25, ferraro_25, ceccarelli_26, petralia_26} aimed at uncovering the origin and nature of the different stellar systems hosted by the MW, placing special focus on faint and observationally challenging clusters. 
On the photometric front, we can take full advantage of {\it HST} large-scale programs, such as the MGCS \citep{massari_25}. Simultaneously, significant advancements are being made regarding the chemical characterization of these systems through the extensive use of archival data and current facilities \citep[e.g.,][]{geisler_21,geisler_25,pace_23,pace_25,garro_26,geha26_I}. 
Combined, all these synergistic efforts will have the distinct advantage of completing our view of the MW's formation history by providing a fully homogeneous characterization of its star cluster populations.

\begin{acknowledgements}
Based on observations collected at the ESO-VLT under the program 0114.D-0288 (P.I. E. Ceccarelli). 
Funded by the European Union (ERC-2022-AdG, "StarDance: the non-canonical evolution of stars in clusters", Grant Agreement 101093572, PI: E. Pancino). Views and opinions expressed are, however, those of the author(s) only and do not necessarily reflect those of the European Union or the European Research Council. Neither the European Union nor the granting authority can be held responsible for them. EL acknowledges support from the ERC Consolidator Grant funding scheme (project ASTEROCHRONOMETRY, G.A. n. 772293).

\end{acknowledgements}
\bibliographystyle{aa} 
\bibliography{biblio}

\end{document}